# Observation of an intrinsic bandgap and Landau level renormalization in graphene/boron-nitride heterostructures

Zhi-Guo Chen[1], Zhiwen Shi[2], Wei Yang[3], Xiaobo Lu[3], You Lai[1], Hugen Yan[4], Feng Wang[2,5*], Guangyu Zhang[3*] and Zhiqiang Li[1*]

[1]National High Magnetic Field Laboratory, Tallahassee, Florida 32310, USA

[2]Department of Physics, University of California at Berkeley, Berkeley, California 94720, USA

[3]Beijing National Laboratory for Condensed Matter Physics and Institute of Physics, Chinese Academy of Sciences, Beijing 100190, China

[4]IBM Thomas J. Watson Research Center, Yorktown Heights, New York 10598, USA

[5]Materials Science Division, Lawrence Berkeley National Laboratory, Berkeley, California 94720, USA

*e-mail: zli@magnet.fsu.edu (Z.L.); fengwang76@berkeley.edu (F.W.); gyzhang@iphy.ac.cn (G.Z.)

**Van der Waals heterostructures formed by assembling different two-dimensional atomic crystals into stacks can lead to many new phenomena and device functionalities. In particular, graphene/boron-nitride heterostructures have emerged as a very promising system for band engineering of graphene. However, the intrinsic value and origin of the bandgap in such heterostructures remain unresolved. Here we report the observation of an intrinsic bandgap in epitaxial graphene/boron-nitride heterostructures with zero crystallographic alignment angle. Magneto-optical spectroscopy provides a direct probe of the Landau level transitions in this system and reveals a bandgap of ~ 38 meV (440 K). Moreover, the Landau level transitions are characterized by effective Fermi velocities with a critical dependence on specific transitions and magnetic field. These findings highlight the important role of many body interactions in determining the fundamental properties of graphene heterostructures.**

Heterostructures consisting of two-dimensional (2D) layers of graphene, hexagonal boron nitride (h-BN), $MoS_2$ and so on coupled by van der Waals interactions[1,2] exhibit many intriguing physical properties[2] and new device functionalities that are not achievable by individual constituting materials[3-7]. In particular, graphene/h-BN heterostructures have shown great potentials for band structure engineering of graphene[8-20] including inducing a bandgap[11,13,15-20] (Fig. 1a), which is of great fundamental[21-24] and technological[25] interest. The coupling between graphene and h-BN results in a periodic moiré superlattice potential due to a 1.8% lattice mismatch[8], which gives rise to superlattice minibands and new Dirac points near the edges of the superlattice Brillouin zone[8-12,14]. Furthermore, the local sublattice symmetry of graphene is broken due to different local potentials produced by boron and nitrogen atoms[17-19] (Fig. 1b), inducing a local bandgap[18,19]. Although this effect varies spatially and is predicted to nearly disappear after spatial averaging[18], transport studies showed signatures of a global bandgap in these heterostructures[11,13]. It is suggested that many body interactions may be responsible for the observed bandgap[15,16], but the issue remains unresolved experimentally.

Here, we report the observation of a finite bandgap in epitaxially grown graphene/h-BN heterostructures with zero crystallographic rotation angle[12] (Fig. 1c, see Methods) by probing their Landau levels (LL) employing magneto-optical spectroscopy. The inter-band (inter-LL) transition peaks measured by optical spectroscopy are determined by peaks in the joint density of states between two bands (LLs), which are not limited by disorder[26,27] such as impurities and defects and therefore enable the measurement of the intrinsic bandgap. On the other hand, disorder may lead to a reduced mobility gap (or thermal activation gap) measured by other techniques compared to the intrinsic gap[26,27].

At zero magnetic field, the energy dispersion of graphene with a bandgap $\Delta$ is[11,28]

$E(p) = \pm\sqrt{v_F^2 p^2 + (\Delta/2)^2}$, where $v_F$ is the Fermi velocity and $p$ is momentum (Fig. 1a). In a magnetic field, $B$, the electronic spectrum of pristine graphene is quantized into LLs described by[11,28]:

$$E_n = sgn(n)\sqrt{2e\hbar v_F^2 B|n|}\,, \qquad (1)$$

where $e$ is the elementary charge, $\hbar$ is Planck's constant divided by $2\pi$, the integer $n$ is LL index, and sgn(n) is the sign function. The LLs for gapped graphene have the form[28]:

$$E_n = \pm(\Delta/2)\delta_{n,0} + sgn(n)\sqrt{2e\hbar v_F^2 B|n| + (\Delta/2)^2}\,, \qquad (2)$$

which features two zeroth LLs labeled as $n = +0$ and $n = -0$ with energies of $E_{\pm 0} = \pm\Delta/2$. Here $\delta$ is the Kronecker delta function. Therefore, the bandgap of graphene can be explored by probing its LL energy spectrum.

Our study provides a direct spectroscopic determination of the bandgap in epitaxial graphene/h-BN heterostructures from optical measurements of LL transitions. We observe an intrinsic bandgap of ∼ 38 meV (440 K) in this system, which is comparable to the gap value found in transport studies[11,13]. Moreover, we find different values of effective Fermi velocity for different LL transitions, indicating LL renormalization by interaction effects. These findings have broad implications for the fundamental understanding of graphene heterostructures and their potential applications.

## Results

**Transmission spectra in magnetic field.** Infrared transmission spectra $T(B)$ were measured in magnetic field applied perpendicular to the samples as shown in Fig. 1d (see Methods). Figure 2 depicts the $T(B)/T(B_0)$ spectra for a representative sample, where $B_0 = 0$ T. Data for more samples are shown in Supplementary Figure 1. Three dip features denoted as $T_1$, $T_2$ and $T_3$ are observed, all of which systematically shift to higher energies with increasing magnetic field. The zero-field transmission spectrum $T(B_0)$ of either pristine or gapped graphene shows a step-like feature without any sharp resonances in the energy range explored here, so the observed dip features in the $T(B)/T(B_0)$ spectra are corresponding to transmission minima in $T(B)$ and thus absorption peaks in magnetic fields (Supplementary Figure 2 and Supplementary Note 1).

The effective bulk mobility of our samples estimated from the widths of the resonances in the optical spectra is higher than 50,000 cm$^2$ V$^{-1}$ s$^{-1}$ (Supplementary Figure 3 and Supplementary Note 2). Our optical data also indicate that the Fermi energy for our samples is in the range of $E_F < 19$ meV (Supplementary Note 3).

**Observed Landau level transitions.** The energies ($E$) of all observed features in graphene/h-BN exhibit an approximate linear dependence on $\sqrt{B}$ (or equivalently, $E^2$ has an approximate linear dependence on $B$) in our spectral range as shown in Fig. 3. However, they all show non-zero energy intercepts at zero magnetic field under linear extrapolations, in stark contrast to the LL transitions of pristine graphene described by equation (1), which converge to zero energy at zero field[29,30]. Similar behaviors were observed in all five samples we have measured (Supplementary Figure 1). Within the non-interacting single particle picture, the finite zero-field extrapolation values of all observed absorption energies suggest that the LLs of our graphene/h-BN samples are described by equation (2). From the selection rule[31] for allowed optical transitions from $LL_n$ to $LL_{n'}$, $\Delta n = |n| - |n'| = \pm 1$, and a quantitative comparison with equation (2), we assign feature $T_1$ to transitions of $LL_{-1} \to LL_{+0}$ and $LL_{-0} \to LL_{+1}$ (Fig. 2b), with an energy given by:

$$E_{T1} = \sqrt{2e\hbar v_F^2 B + (\Delta/2)^2} + \Delta/2 \qquad (3)$$

Fitting the $T_1$ feature based on equation (3) from a least squares fit yields a bandgap $\Delta \approx 38 \pm 4$ meV and an effective Fermi velocity $v_F^{T1} \approx (0.96 \pm 0.02) \times 10^6$ m s$^{-1}$ (Supplementary Table 1 and Supplementary Note 4). We emphasize that the bandgap explored here is at the main Dirac point of graphene instead of the secondary Dirac points at the edges of graphene/BN superlattice Brillouin zone[8-12]. The $T_1$ transition energies in Fig. 2 are well below the energy (~ 200 meV) of the secondary Dirac points[9-12], so the LLs in this energy region are not significantly affected by the strong band structure modifications at the boundary of the superlattice Brillouin zone, which is supported by the observed linear $\sqrt{B}$ dependence of the LL transition energy.

The observed $T_2$ and $T_3$ features have higher energies compared to $T_1$ features and can be assigned as (Fig. 2b): $T_2$, $LL_{-2} \to LL_{+1}$ and $LL_{-1} \to LL_{+2}$; $T_3$, $LL_{-3} \to LL_{+2}$ and $LL_{-2} \to LL_{+3}$. Their energies are described by:

$$E_{T2} = \sqrt{2e\hbar v_F^2 B + (\Delta/2)^2} + \sqrt{(2e\hbar v_F^2 B) \times 2 + (\Delta/2)^2} \qquad (4)$$

$$E_{T3} = \sqrt{(2e\hbar v_F^2 B) \times 2 + (\Delta/2)^2} + \sqrt{(2e\hbar v_F^2 B) \times 3 + (\Delta/2)^2} \qquad (5)$$

The energies of $T_2$ transition show a deviation from linear $\sqrt{B}$ dependence above 4 T (or 220 meV in energy) as shown in Fig. 3, with $E^2$ deviating from a linear $B$-dependence, which is perhaps due to the effect of moiré superlattice or many body interactions. Therefore, we focus on the low field (< 4 T) region where the $T_2$ transition exhibits an overall linear $\sqrt{B}$ dependence, which most likely arises from the intrinsic behaviors of gapped graphene alone. We observed a splitting of the $T_2$ transition near 169 meV due to the coupling to the infrared active phonon of h-BN, which nonetheless doesn't affect the main conclusions of our analysis because this effect only occurs in a very narrow field and energy range. Based on equation (4), we find that the $T_2$ transition in low field is consistent with a bandgap similar to that extracted from the $T_1$ transition, $\Delta \approx 38 \pm 4$ meV, and an effective Fermi velocity $v_F^{T2} \approx (1.20 \pm 0.01) \times 10^6$ m s$^{-1}$ (Supplementary Figure 4 and Supplementary Note 4). The $T_3$ transition is discussed in details below and in Supplementary Figure 5 and Supplementary Note 4.

**Comparison with pristine and gapped graphene.** We stress that our data on graphene/h-BN cannot be explained by many body effects of pristine graphene (Supplementary Figure 6 and Supplementary Note 5). One prominent feature of interaction effects in pristine graphene is that the effective Fermi velocity varies for different LL transitions, so that the energy ratios $E_{T2}/[(\sqrt{2}+1)E_{T1}]$ and $E_{T3}/[(\sqrt{3}+\sqrt{2})E_{T1}]$ are higher than one, as demonstrated by previous infrared studies[29] and our data on graphene on SiO$_2$. However, the data for graphene/h-BN exhibit an entirely different behavior (Fig. 4a) compared to interaction effects in pristine graphene. Instead, the energy ratios of different LL transitions for graphene/h-BN are consistent with the behaviors of gapped graphene. A theoretical result of $E_{T2}/[(\sqrt{2}+1)E_{T1}]$ based on equations (3) and (4) is shown in Fig. 4a, with $\Delta \approx 38$ meV, $v_F^{T1} = 0.96 \times 10^6$ m s$^{-1}$ and $v_F^{T2} = 1.20 \times 10^6$ m s$^{-1}$, which agrees very well with the experimental results.

# Discussion

Previous transport measurements on graphene/h-BN heterostructures indicated a gap of ~ 300K at 0.4° crystallographic rotational angle $(\theta)$[11]. A recent study[13] reported the existence of large domains of graphene with the same lattice constant as hBN separated by domain walls with concentrated strain for small $\theta$, and a gap of 360 K was found for $\theta = 0°$. Our optical study provides a direct spectroscopic determination of the bandgap with similar magnitude (~ 440 K) in epitaxial graphene/h-BN heterostructures. This bandgap value is larger than those found in theories within the single particle picture[18-20], which suggests the relevance of many body interactions in generating the gap[15,16]. It was argued that the gap at the Dirac point is greatly enhanced by interaction effects due to coupling to a constant sublattice-asymmetric superlattice potential[15], which is not affected by the spatial variations shown in Fig. 1b. The intrinsic gap value for graphene/h-BN obtained in our study provides a critical input for the basic understanding of the gap in this system.

Our study further reveals the crucial role of many-body interactions in renormalizing LL transitions[32] in graphene/h-BN heterostructures. Specifically, the effective Fermi velocity associated with the LL transitions varies with particular transitions as well as the magnetic field. We find that the $T_3$ transition cannot be consistently fitted by equation (5) using a constant $v_F$, so we employ an *effective* field-dependent parameter $v_F^{T3}(B)$ to describe this transition (Supplementary Note 4). Fig. 4b depicts the Fermi velocity ratios $v_F^{T2}/v_F^{T1}$ and $v_F^{T3}(B)/v_F^{T1}$, both of which are higher than one and therefore very different from the constant $v_F$ for all transitions expected from single particle pictures. Intriguingly, $v_F^{T3}(B)/v_F^{T1}$ shows a systematic increase in low magnetic fields. For $T_2$ transitions (consider $LL_{-1} \rightarrow LL_{+2}$ for example), it shows $v_F^{T2} \sim 1.20 \times 10^6$ m s$^{-1}$ even at 1 Tesla field with $E_{T2} = 115$ meV (Supplementary Figure 1c), which corresponds to $E_{LL+2} \sim 66$ meV and $E_{LL-1} \sim 49$ meV. According to theoretical studies[33], the band structure of graphene/BN at such low energy scales are quite linear and not strongly modified by the superlattice Dirac points ($\sim 200$ meV[9-12]), so the value $v_F^{T2} \sim 1.20 \times 10^6$ m s$^{-1}$ extracted from data at low magnetic field (thus low energy) is little affected by the superlattice Dirac points. Similar argument can be made for $T_1$ and $T_3$ transitions at low magnetic fields and low energy. Note that the LL transitions at high field and high energy (for instance, $T_2$ transition above 4T field) may be affected by the band structure modification due to the superlattice Dirac points[8, 33], but our discussions here are only focused on the low field regime shown in Fig. 4b. Our results in Fig. 4b indicate LL renormalization due to many-body interactions in magnetic field. Theoretical studies[34-37] showed that interaction effects of electron-hole excitations between LLs, such as direct Coulomb interactions between the excited electrons and holes and the exchange self-energy of electrons and holes between LLs, can significantly renormalize the inter-LL transition energy. The observation shown in Fig. 4b in gapped graphene is qualitatively similar to the results from many body theories[34-37] as well as experimental studies[29] on pristine graphene, so our results strongly suggest contributions of many body effects to the inter-LL transitions[34-37]. Further theoretical investigations are required to quantitatively understand these interactions in gapped graphene, with many open questions yet to be addressed such as the role of superlattice potential[15] and bond distortion[20] in graphene/h-BN heterostructures.

Multi-valley (band extrema in momentum space) Dirac systems such as gapped graphene, silicene and 2D transition metal dichalcogenides are described by the same Dirac Hamiltonian and share several essential properties such as valley-dependent orbital magnetic moment and Berry curvature[23,38], which are intimately related to their unconventional valley-dependent LL structures[38-40]. In this context, the strong LL renormalization observed here has broad implications for fundamental studies of many novel phenomena related to LLs in these Dirac materials, such as magnetic control of valley degree of freedom[38] and valley-spin polarized magneto-optical response[39,40].

In summary, we have observed a bandgap of ~ 38 meV (440 K) in graphene/h-BN heterostructures with zero crystallographic rotation angle employing magneto-optical spectroscopy. The intrinsic gap value reported here is important for fundamental understanding of the bandgap and many body interaction effects in this system. Our demonstration of a finite bandgap in epitaxial graphene/h-BN heterostructures can also lead to novel applications in electronics and optoelectronics.

# Methods

**Sample preparation and characterization**. Hexagonal boron nitride (h-BN) was mechanically exfoliated onto double-side-polished $SiO_2/Si/SiO_2$ substrates with 300 nm $SiO_2$. Graphene was epitaxially grown on h-BN by remote plasma enhanced chemical vapor deposition[12]. Some multilayer grains can be found on monolayer graphene in as-grown samples, so hydrogen plasma etching technique[12] was applied to reduce these additional grains. The resulting sample is continuous monolayer graphene with minor etched hexagonal pitches in plane, as shown in Fig. 1c. The moiré pattern (Fig. 1c) due to lattice mismatch shows a periodicity of 15 ± 1 nm as measured by atomic force microscopy (AFM), indicating zero crystallographic alignment angle between graphene and h-BN[12]. This moiré pattern is observed over the entire areas of all samples, establishing these epitaxial samples as single-crystalline and single-domain graphene heterostructures. The samples studied in this work have typical lateral sizes of about 100 microns. The observed sharp Landau level transitions indicate that the optical absorption of our samples is little affected by defects or grain boundaries. The effective mobility of our samples estimated from the widths of the resonances in the optical spectra is higher than 50,000 $cm^2\,V^{-1}\,s^{-1}$ (Supplementary Figure 3 and Supplementary Note 2). The absence of the $LL_{-1} \rightarrow LL_{-0}$ and $LL_{+0} \rightarrow LL_{+1}$ transitions in our optical data indicates that the Fermi energy is within the gap for our samples, namely $E_F < 19$ meV (Supplementary Note 3).

**Magneto-transmission measurements.** The measurements were performed at ~ 4.5 K in a superconducting or resistive magnet in the Faraday geometry (magnetic field perpendicular to the sample surface). Infrared light from a Fourier transform spectrometer is delivered to the sample using a copper light pipe, and the light transmitted through the sample is detected by a composite Si bolometer. The focus of the IR light on the sample is about 0.5-1 mm. To reduce the stray light around our small samples, an aluminum aperture ~ 200 microns diameter was placed around the sample. We report data at energies above 60 meV corresponding to wavelengths shorter than 20 microns, which ensures that the wavelength is significantly smaller than the sizes of the samples and therefore a macroscopic description of the data using optical constants is applicable.

## Acknowledgements

Z.C., Y.L. and Z.L. acknowledge support from the UCGP program at NHMFL. Z.S. and F.W. are supported by Office of Basic Energy Science, Department of Energy under contract No. DE-SC0003949 (Early Career Award). G.Z. acknowledges the supports from the National Basic Research Program of China (973 Program, grant No. 2013CB934500), the National Science Foundation of China (NSFC, grant Nos. 61325021, 91223204), and the Chinese Academy of Sciences. Optical measurements were performed at the National High Magnetic Field Laboratory, which is supported by National Science Foundation Cooperative Agreement No. DMR-1157490, the State of Florida, and the U.S. Department of Energy.

## Author Contributions

Z.C. and Z.S. initiated the optical studies. Z.C. carried out the optical experiments. Z.S. and Y.L. participated in part of the measurements. W.Y. and X.L. grew and characterized the graphene/h-BN samples. H.Y. provided the graphene on $SiO_2$ samples. F.W., G.Z. and Z.L. supervised the project. Z.C. and Z.L. analyzed the data and wrote the manuscript. All authors discussed the results and commented on the paper.

## Competing Financial Interests

The authors declare no competing financial interests.

# Figures and Figure Legends

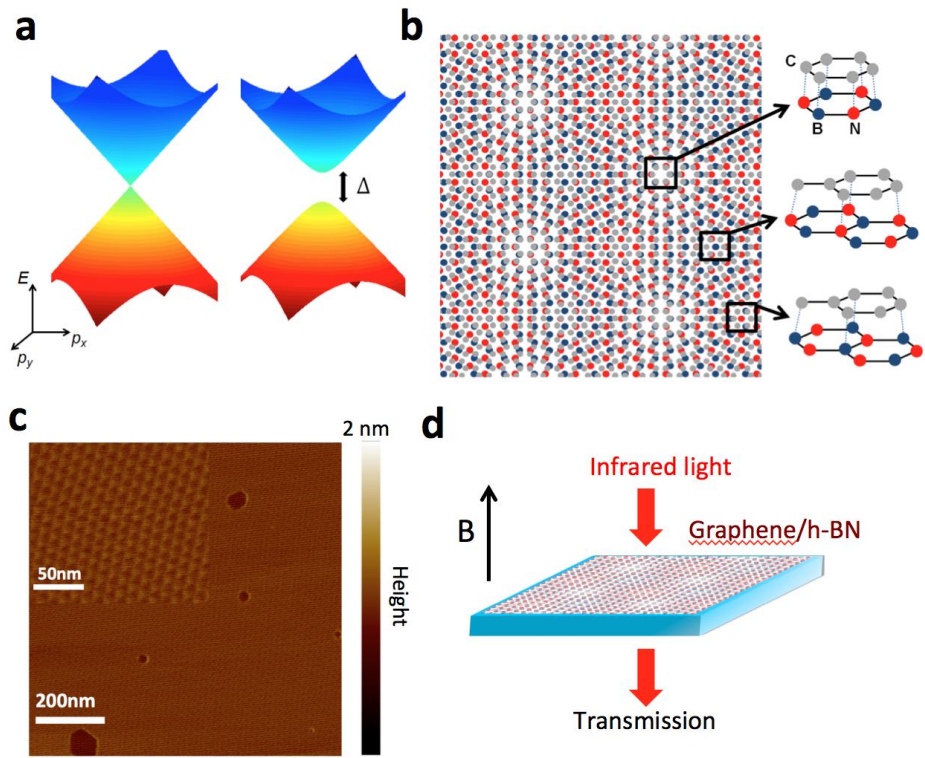

**Figure 1 | Graphene/h-BN heterostructures. (a)** Energy spectrum of pristine graphene (left) and gapped graphene (right). **(b)** Schematic of the moire pattern in graphene on h-BN with zero crystallographic rotation angle and an exaggerated lattice mismatch of 11% (carbon, gray; boron, blue; nitrogen, red). The lattice alignments in different regions lead to different local sublattice symmetry breaking in graphene. **(c)** AFM image of a monolayer graphene sample grown on h-BN and treated by hydrogen plasma etching, with bare BN shown in dark colour. The inset shows the observed moiré pattern with a periodicity of 15 ± 1 nm. **(d)** Schematic of the magneto-optical measurements.

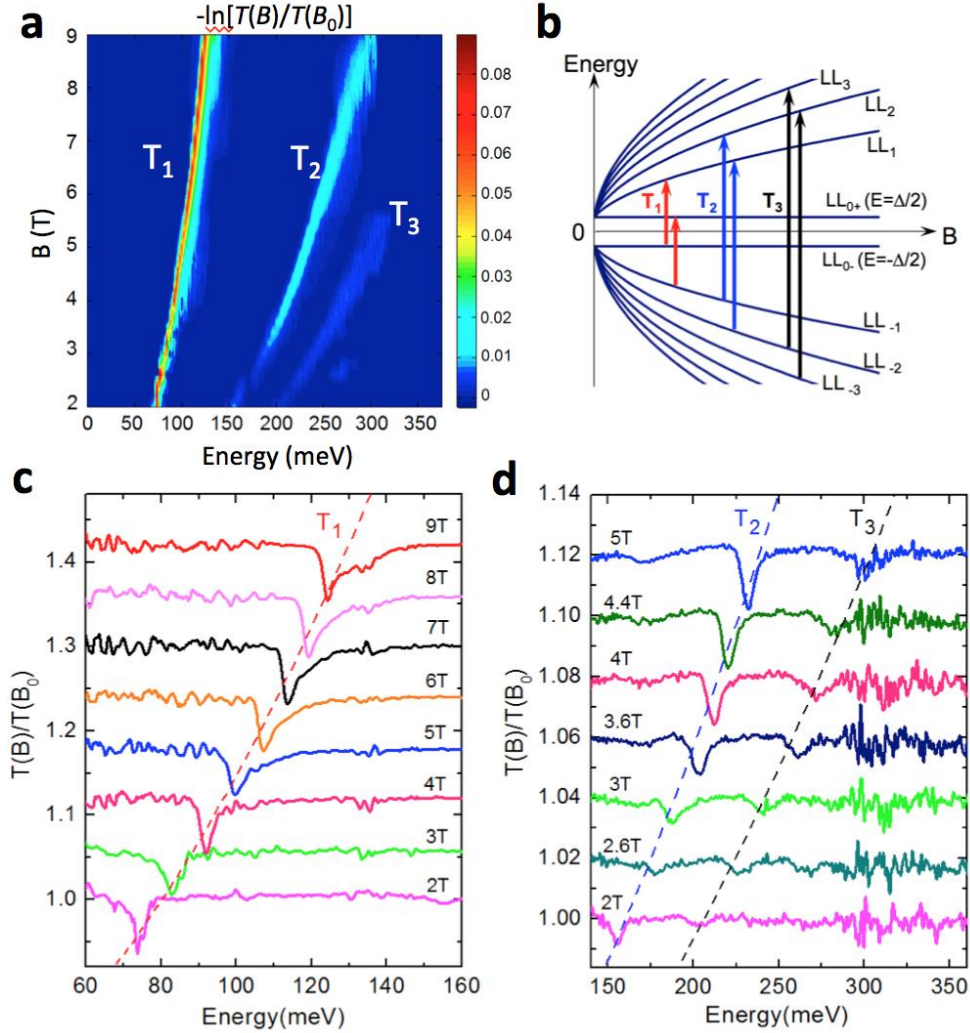

**Figure 2 | Magneto-transmission ratio spectra of graphene/h-BN. (a)** Colour rendition of the $-\ln[T(B)/T(B_0)]$ spectra as a function of magnetic field and energy for sample 1, where $B_0 = 0$ T. **(b)** Schematic of Landau levels of gapped graphene. The arrows indicate transitions observed in this study. **(c,d)** Several representative $T(B)/T(B_0)$ spectra for sample 1; dashed lines are guides to eyes. For clarity the data in panels **c** and **d** are displaced from one another by 0.06 and 0.02, respectively.

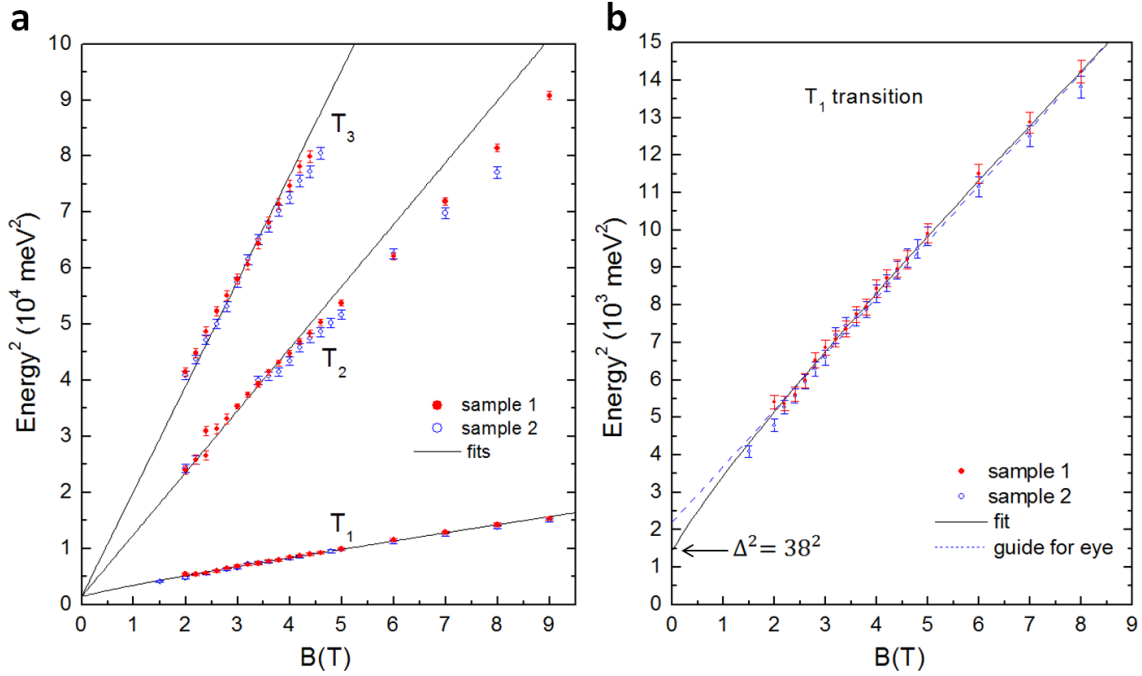

**Figure 3 | Landau level transition energies of graphene/h-BN. (a)** All observed transitions shown in a $E^2$-$B$ plot. Symbols: data for sample 1 and 2. Solid lines: best fits to the data for sample 1 using equations (3)-(5) and parameters discussed in the text. $\Delta = 38$ meV and $v_F^{T3} \approx 1.20 \times 10^6$ m s$^{-1}$ are used for the fit to T$_3$ transition shown here. **(b)** The low energy part of panel **a** to highlight the extraction of the gap. Dashed line: a guide for eye showing linear extrapolation of the data. The error bars in both panels, $\delta(E^2)$, are calculated as $\delta(E^2) = 2E\delta(E)$, where $\delta(E)$ is the uncertainty in determining the energy of each Landau level transition from the $T(B)/T(B_0)$ spectra.

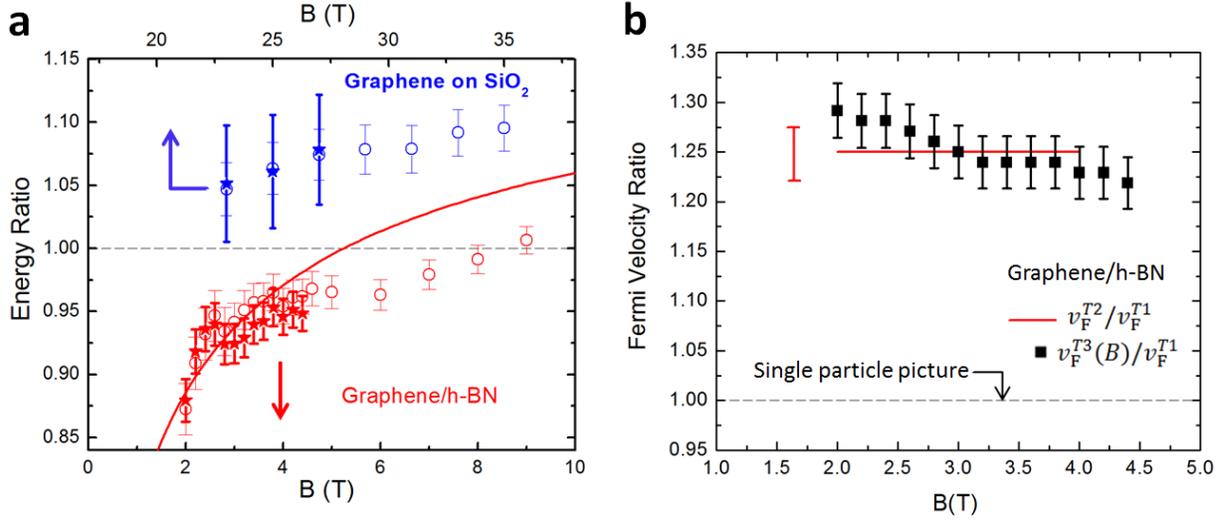

**Figure 4 | Many body effects on Landau level transitions for pristine and gapped graphene. (a)** Energy ratios of different Landau level transitions for graphene on SiO$_2$ (blue colour) and graphene/h-BN sample 1 (red colour) shown in a common vertical scale. Open symbols: $E_{T2}/[(\sqrt{2}+1)E_{T1}]$. Solid symbols: $E_{T3}/[(\sqrt{3}+\sqrt{2})E_{T1}]$. Red solid line: theoretical result of $E_{T2}/[(\sqrt{2}+1)E_{T1}]$ based on equations (3) and (4) for gapped graphene. The ratios for graphene on SiO$_2$ are greater than one, which is a signature of interaction effects in pristine graphene (Supplementary Note 5). On the other hand, the ratios for graphene/h-BN exhibit an entirely different behavior, which is consistent with gapped graphene. The error bars of energy ratios are calculated using standard formulas for propagation of uncertainty for division based on the uncertainty in determining the energy of each Landau level transition from the $T(B)/T(B_0)$ spectra. **(b)** Fermi velocity ratios of different Landau level transitions for graphene/h-BN with a constant $v_F^{T1} \approx (0.96 \pm 0.02) \times 10^6$ m s$^{-1}$. For T$_2$ transition, a constant Fermi velocity $v_F^{T2} \approx (1.20 \pm 0.01) \times 10^6$ m s$^{-1}$ is extracted from the data. These ratios are distinct from the value expected from the single particle picture, which is indicative of interaction effects in gapped graphene. The error bars indicate the range of Fermi velocity values that could fit the data in Fig. 3 (Supplementary Note 4).

# Supplementary Information

Zhi-Guo Chen[1], Zhiwen Shi[2], Wei Yang[3], Xiaobo Lu[3], You Lai[1], Hugen Yan[4], Feng Wang[2,5*], Guangyu Zhang[3*] and Zhiqiang Li[1*]

[1]National High Magnetic Field Laboratory, Tallahassee, Florida 32310, USA

[2]Department of Physics, University of California at Berkeley, Berkeley, California 94720, USA

[3]Beijing National Laboratory for Condensed Matter Physics and Institute of Physics, Chinese Academy of Sciences, Beijing 100190, China

[4]IBM Thomas J. Watson Research Center, Yorktown Heights, New York 10598, USA

[5]Materials Science Division, Lawrence Berkeley National Laboratory, Berkeley, California 94720, USA

*e-mail: zli@magnet.fsu.edu (Z.L.); fengwang76@berkeley.edu (F.W.); gyzhang@iphy.ac.cn (G.Z.)

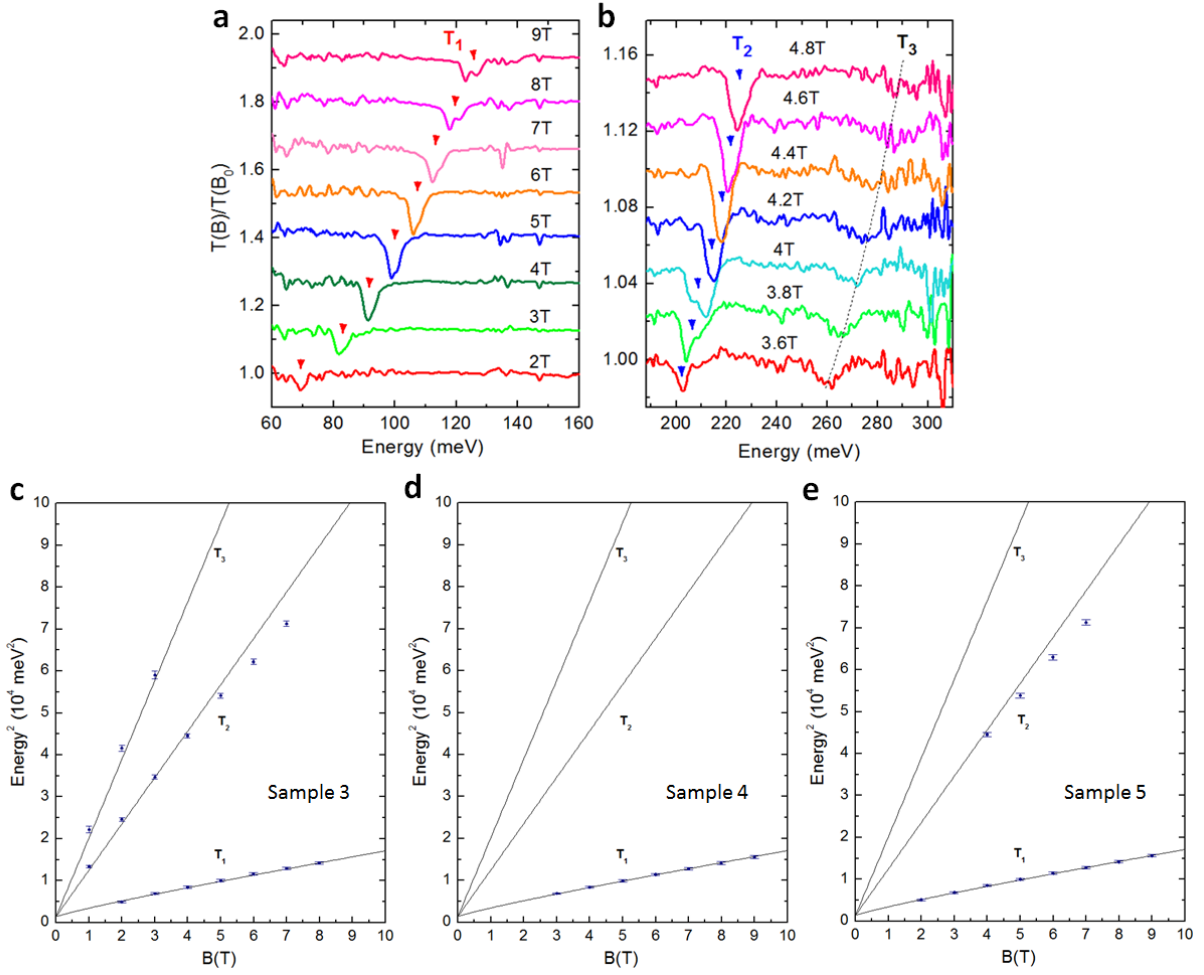

**Supplementary Figure 1 | Magneto-transmission spectra of graphene/h-BN sample 2 and Landau level transition energies of three other samples. (a,b)** Magneto-transmission ratio spectra $T(B)/T(B_0)$ of graphene/h-BN sample 2 in representative magnetic fields, where $B_0 = 0$ T. For clarity the data in panels **a** and **b** are displaced from one another by 0.133 and 0.025, respectively. Coloured triangles and the dashed line are guides to eyes. **(c-e)** The observed transition energies presented in a $E^2$-$B$ plot for graphene/h-BN samples 3, 4 and 5. The error bars for the $T_1$ transition data in panels **c-e** are similar to the size of the symbols. Solid lines in panels **c-e** are theoretical results identical to the fits displayed in figure 3 of the main text. The error bars in panels c-e, $\delta(E^2)$, are calculated as $\delta(E^2) = 2E\delta(E)$, where $\delta(E)$ is the uncertainty in determining the energy of each Landau level transition from the $T(B)/T(B_0)$ spectra.

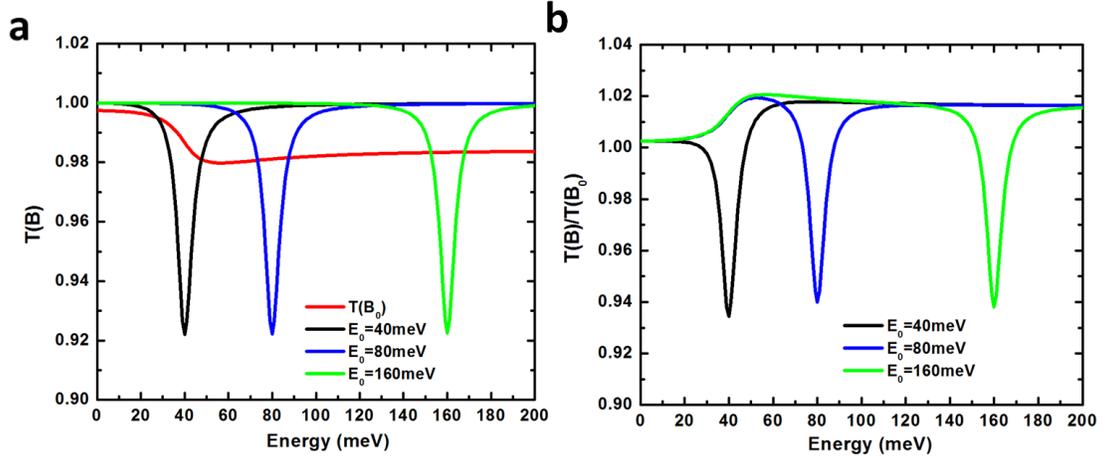

**Supplementary Figure 2 | Simulations of $T(B_0)$ and $T(B)/T(B_0)$ spectra.** **(a)** a simulation of the zero field transmission spectrum $T(B_0)$ of gapped graphene with a gap of 40 meV ($B_0$=0T), and several representative $T(B)$ spectra of LL transitions simulated by Lorentzians with different resonance energy $E_0$. **(b)** $T(B)/T(B_0)$ spectra calculated from panel (a).

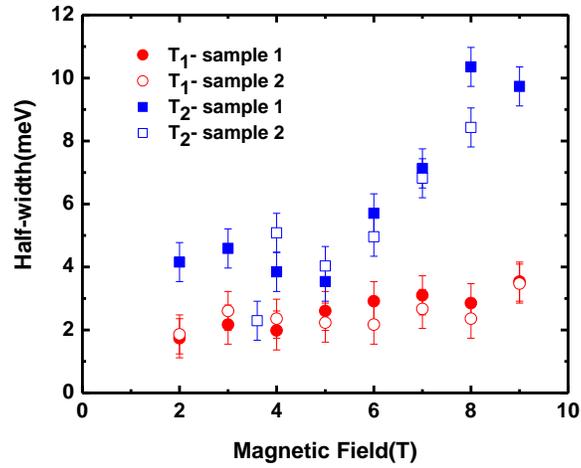

**Supplementary Figure 3 | The widths of the $T_1$ and $T_2$ transitions for graphene/h-BN samples 1 and 2.** The error bars correspond to the uncertainty in determining the width of each Landau level transition from fitting the $T(B)/T(B_0)$ spectra with Lorentzian oscillators.

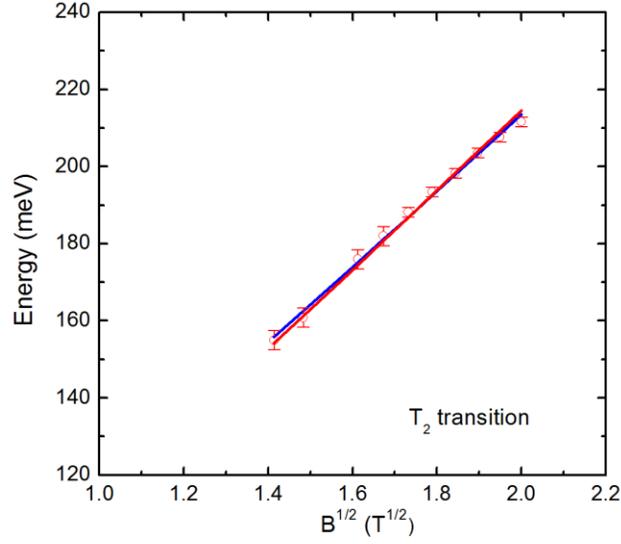

**Supplementary Figure 4 | Least squares fit for the $T_2$ transition in magnetic fields below 4T.**
Symbols: experimental data. Blue curve: best fit from the first approach with $\Delta' \approx 54 \pm 6$ meV and $v_F^{T2'} \approx (1.18 \pm 0.01) \times 10^6$ m s$^{-1}$, with $\chi^2 = 1.7$. Red curve: best fit from the second approach with $\Delta \approx 38 \pm 4$ meV and $v_F^{T2} \approx (1.20 \pm 0.01) \times 10^6$ m s$^{-1}$, with $\chi^2 = 2.3$. The two fits are equally satisfactory. The error bars correspond to the uncertainty in determining the energy of the Landau level transition from the $T(B)/T(B_0)$ spectra.

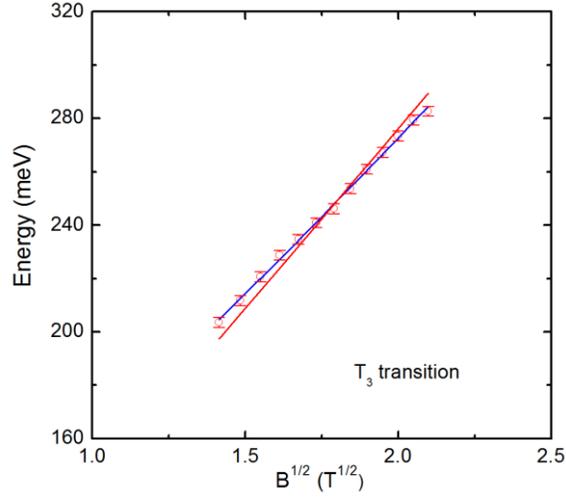

**Supplementary Figure 5 | Least squares fit for the $T_3$ transition.** Symbols: experimental data. Blue curve: best fit from the first approach with $\Delta' = 96 \pm 3$ meV and $v_F^{T3'} \approx (1.12 \pm 0.01) \times 10^6$ m s$^{-1}$, with $\chi_1^2 = 1.1$. Red curve: best fit from the second approach with $\Delta \approx 38 \pm 4$ meV and $v_F^{T3} \approx (1.20 \pm 0.01) \times 10^6$ m s$^{-1}$, with $\chi_2^2 = 16.5$. It is clear that the data cannot be satisfactorily described by the second approach (red curve) with a gap value $\Delta \approx 38 \pm 4$ meV. The error bars correspond to the uncertainty in determining the energy of the Landau level transition from the $T(B)/T(B_0)$ spectra.

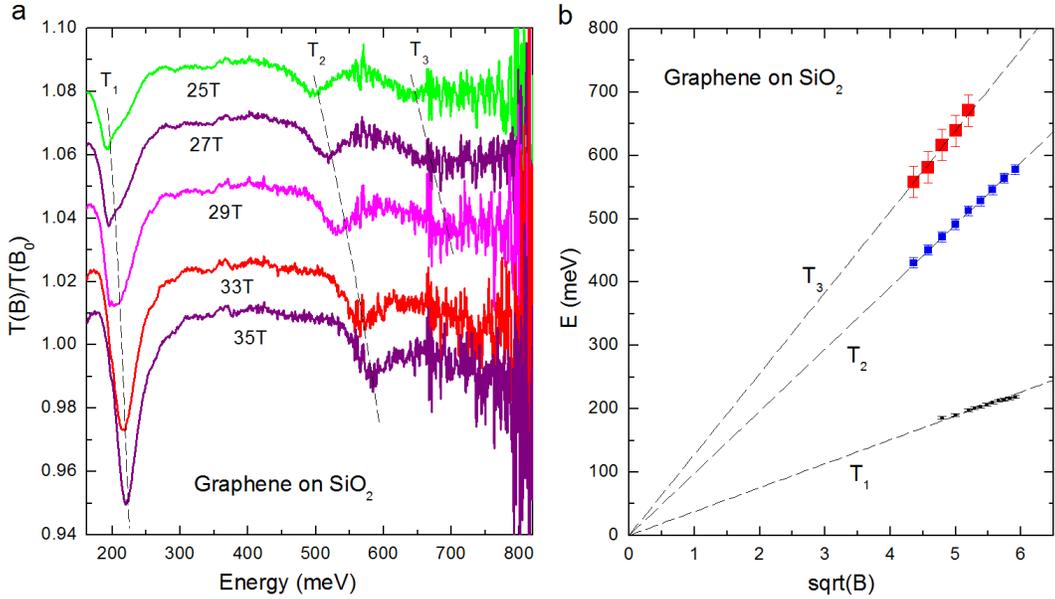

**Supplementary Figure 6 | Magneto-transmission spectra and Landau level transition energies of CVD graphene on SiO$_2$.** (a) magneto-transmission ratio spectra of CVD graphene on SiO$_2$. Data are displaced by 0.02 from one another for clarity. Dashed lines are guides to eyes. (b) Landau level transition energies as a function of $\sqrt{B}$. Dashed lines are fits using equation (1) in the main text. The error bars correspond to the uncertainty in determining the energy of each Landau level transition from the $T(B)/T(B_0)$ spectra.

|                | Gap $\Delta$ (meV) | $v_F^{T1}$ ($10^6$ m s$^{-1}$) |
|----------------|--------------------|--------------------------------|
| Sample 1 (z25) | 40.38±0.99         | 0.945±0.008                    |
| Sample 2 (z2)  | 36.72±1.14         | 0.960±0.010                    |
| Sample 3 (n25) | 35.63±2.11         | 0.984±0.016                    |
| All 5 samples  | 37.49±0.78         | 0.962±0.006                    |

**Supplementary Table 1 | The value and uncertainty of $\Delta$ and $v_F^{T1}$ extracted from fitting the $T_1$ transition of graphene/h-BN samples.**

**Supplementary Note 1 | Zero-field transmission spectra and analysis of *T*(*B*)/*T*(*B*₀) data.**

Here we examine features in the zero-field transmission spectrum of pristine graphene and gapped graphene, which can elucidate the origin of the sharp dip features in the *T*(*B*)/*T*(*B*₀) spectra of graphene on h-BN, where $B_0$ = 0 T. The real part of the optical conductivity spectrum $\sigma_1(\omega)$ (corresponding to absorption) of pristine graphene shows a Drude absorption peak at zero energy and a step-like feature at $2E_F$, above which the conductivity spectrum is a constant in the infrared range[1]. Here, we discuss the case of finite $E_F$ relevant to real samples, which always have electron and hole puddles induced by impurities and thus finite $E_F$. While $\sigma_1(\omega)$ spectrum of gapped graphene featuring massive Dirac fermions has not been reported experimentally yet, theoretical calculations showed that $\sigma_1(\omega)$ is characterized by a Drude peak and a step-like feature at the higher energy of $2E_F$ or band gap Δ due to the onset of interband transitions[2], above which $\sigma_1(\omega)$ spectrum has a form of $\frac{\omega^2+\Delta^2}{\omega^2}$. (Note that we define Δ as the full band gap in this paper.)

The half-width of the observed absorption features in graphene/h-BN is about 3-10 meV, which is roughly corresponding to the scattering rate. Therefore, the Drude component is narrow (~ 3-10 meV) and not in the energy range discussed in this work. To simulate the interband transition contributions, we calculate the zero field transmission *T*(*B*₀) based on the $\sigma_1(\omega)$ spectrum of gapped graphene at zero field (equation (13) in Supplementary Reference 3), with a scattering rate of 4 meV, an energy gap of 40 meV, a temperature 4.5 K and a chemical potential smaller than half of the energy gap, all of which are based on real parameters in our experiments. The *T*(*B*₀) spectrum shows a step-like feature without any sharp resonance peaks (Supplementary Figure 2a). The transmission *T*(*B*) due to Landau level (LL) transitions in magnetic field can be simulated using Lorentzians with a half-width of 4 meV and with several resonance energies $E_0$ to illustrate the effect of different locations of the resonance with respect to the step-like feature in *T*(*B*₀). The transmission ratio *T*(*B*)/ *T*(*B*₀) from these simulations are shown in Supplementary Figure 2b, which shows that the sharp dip features in *T*(*B*)/ *T*(*B*₀) are all due to sharp minima in *T*(*B*) in different cases, because of the lack of sharp features in the *T*(*B*₀) spectrum. Similar results can be found from an analysis of pristine graphene that has a constant $\sigma_1(\omega)$ above $2E_F$.

For simplicity, substrate effects are neglected in the discussions above. The presence of the $SiO_2$/Si substrate, specifically the Fabry–Pérot interference due to the 300 nm $SiO_2$ layer, gives rise to a modulation to the lineshape of the overall transmission spectrum of the samples. This lineshape modulation is very broad with a half-width of ~ 700 meV (Supplementary Reference 4), which can be neglected compared to the observed sharp resonances (half-width ~ 5 meV) observed in our graphene/h-BN samples. The Fabry–Pérot interference effects due to h-BN can be neglected as well, because the h-BN layer in our samples is typically much thinner than 300nm, which will lead to even broader lineshape modulation, because the width of the Fabry–Pérot interference is inversely proportional to the thickness of the layer.

**Supplementary Note 2 | Mobility and bandgap of epitaxial graphene/h-BN samples from transport and optical experiments.**

Supplementary Figure 3 shows the width of the LL transitions as a function of magnetic field for graphene/h-BN sample 1 and 2. The widths of the $T_2$ and $T_3$ transitions are broader than that of the $T_1$ transition.

The widths of the absorption peaks due to LL transitions are mainly determined by the scattering rate of charge carriers in low-mobility samples dominated by disorder. Remarkably, the widths of the LL transition features for graphene/h-BN are substantially (~ 10 times) narrower compared to those of exfoliated graphene deposited on $SiO_2$ (Supplementary Reference 5), which indicates much lower scattering rate in graphene/h-BN. Because the mobility is inversely proportional to the scattering rate, the half width of the $T_1$ transition (~ 2-4 meV) suggests an effective mobility of higher than 50,000 $cm^2\,V^{-1}\,s^{-1}$ for our graphene/h-BN samples, which is more than 10 times higher than the value for graphene on $SiO_2$ with mobility ~ 4,000 $cm^2\,V^{-1}\,s^{-1}$ (Supplementary Reference 5) and similar to that reported in graphene on SiC with mobility ~ 50,000 $cm^2\,V^{-1}\,s^{-1}$ (Supplementary Reference 6) obtained from magneto-optical measurements.

Our monolayer epitaxial samples have a point-defect density of about 10 $\mu m^{-2}$ (Supplementary Reference 7), so they have much higher disorder compared to exfoliated graphene samples that are transferred onto h-BN. Due to the presence of nm-scale pits (areas without graphene) in our samples as shown in figure 1(c) of the main text, monolayer graphene has a domain size of several hundred nanometers. The edges and domain boundaries (due to the pits) and point defects in our samples severely affect the electrical transport, limiting the electronic mobility to ~ 5,000 $cm^2\,V^{-1}\,s^{-1}$ in transport measurements[7]. At the charge neutral point, edge transport and defects in these samples can dominate the conductivity, making it very difficult to observe the intrinsic bandgap from transport measurements. On the other hand, optical measurements of LL transitions are less sensitive to the finite domain size, because the cyclotron radius (around tens of nanometers in our magnetic field range) is much less than the domain size of the samples and the cyclotron orbitals of most carriers are not affected by the domain boundaries. Moreover, LL transition peaks measured by optical spectroscopy are not limited by disorder such as defects, as discussed in our manuscript. As a result, we observe very sharp LL transitions in the optical spectra, the width of which corresponds to an effective mobility > 50,000 $cm^2\,V^{-1}\,s^{-1}$. Similarly, optical measurements have allowed us to determine the intrinsic bandgap in these epitaxial samples.

**Supplementary Note 3 | Fermi energy $E_F$ estimated from the observed Landau level transitions.**

From DC transport measurements on gated devices made from our graphene/h-BN samples, we find that the gate dependent resistance peak associated with the original Dirac point is usually observed at gate voltage $|V_g| < 5\,V$, whereas the resistance peak due to secondary Dirac points at the edges of superlattice Brillouin zone[7-10] always appears at very high gate voltage (~ −40 V). These results confirm that the Fermi energy is located near the original Dirac points of graphene at zero gate voltage.

Fermi energy can be estimated from our optical data. Based on Pauli's exclusion principle, the $T_1$ transition will be blocked when the $LL_1$ (or $LL_{-1}$) becomes fully occupied (or depleted) below a critical magnetic field $B_{T1}$, in which $LL_1$ (or $LL_{-1}$) coincides with $E_F$, namely $E_\text{F} = E_{\pm 1}(B_{\text{T}1}) = \sqrt{2e\hbar v_\text{F}^2 B_{\text{T}1} + (\Delta/2)^2}$. As shown in the main text, the T$_1$ transition still has considerable spectral weight at 2 T field for sample 1, which indicates $B_{\text{T}1}$ is well below 2 T. (The T$_1$ transition shifts out of our spectral range below 2 T field, so we cannot determine the value of $B_{\text{T}1}$.) Therefore, the T$_1$ transition data suggest that E$_\text{F}$ is much lower than 52 meV, which is obtained from the above formula using $v_\text{F}^{\text{T}1}$ and $\Delta$ for sample 1 discussed in the main text and $B_{\text{T}1}$ = 2 T.

Moreover, the transitions of LL$_{-1} \rightarrow$ LL$_{-0}$ and LL$_{+0} \rightarrow$ LL$_{+1}$ with energy of $\sqrt{2e\hbar v_\text{F}^2 B + (\Delta/2)^2} - \Delta/2$ were not observed in our measurements. These transitions extrapolate to zero energy at zero magnetic field, which is clearly different from all the observed transitions shown in the main text. The absence of LL$_{-1} \rightarrow$ LL$_{-0}$ and LL$_{+0} \rightarrow$ LL$_{+1}$ transitions in our data suggests that $E_\text{F}$ is within the gap[3] (between LL$_{-0}$ and LL$_{+0}$), so these transitions are not allowed due to Pauli's exclusion principle. This indicates $E_\text{F} < \Delta/2 \sim 19$ meV for our samples.

The carrier density and Fermi energy are spatially inhomogeneous across the sample due to disorder[11], so the $E_\text{F}$ value estimated from our optical measurements is a spatially averaged Fermi energy. The graphene/h-BN samples studied in our optical experiments are as-grown ones, so no additional disorder was induced to the samples by device fabrication processes. The low Fermi energy found in our graphene/h-BN samples is consistent with the high quality of single crystal h-BN, which is free of dangling bonds and charge traps and therefore leads to very low unintentional dopings to the graphene samples[11].

**Supplementary Note 4 | Extracting band gap and effective Fermi velocity from the Landau level transition energies.**

We fit the observed transition energies of graphene/h-BN shown in figure 3 of the main text based on the LL energy of gapped graphene and equations (3)-(5) for $E_{\text{T}1}$, $E_{\text{T}2}$ and $E_{\text{T}3}$ in the main text. We fit the T$_1$ transition data based on equation (3) in the main text using the method of least squares fit, with $\Delta$ and $v_\text{F}$ as free parameters that are field-independent. The Supplementary Table 1 summarizes the value and uncertainty of $\Delta$ and $v_\text{F}^{\text{T}1}$ from fitting the T$_1$ transition. We obtained fewer data points from samples 4 and 5, so they are not fit individually. The analysis of T$_1$ transition data shows a gap $\Delta \approx 38 \pm 4$ meV and an effective Fermi velocity $v_\text{F}^{\text{T}1} \approx (0.96 \pm 0.02) \times 10^6$ m s$^{-1}$, considering all samples in Supplementary Table 1.

As discussed in the main text, we focus on the low field (< 4 T) region for the T$_2$ transition, in which an overall linear $\sqrt{B}$ dependence is observed. The data at 2.4T field is not included in the fit due to the magnetophonon resonance discussed above. In the least squares fit for the T$_2$ transition based on equation (4) in the main text, we used two approaches: (1) allowing both $\Delta$ and $v_\text{F}$ to be free parameters; (2) fixing the gap value to $\Delta = 38 \pm 4$ meV and allowing $v_\text{F}$ to be a free parameter. Both $\Delta$ and $v_\text{F}$ are field-independent in the fit. These two approaches yield quite similar fits within the error bars of our data and comparable values of $\chi^2$, as shown in Supplementary Figure 4. Here $\chi^2$ is the sum of squared residuals, which are the differences between the observed value and the fitted value in the least squares fit. Since $\chi^2$ is indicative of the quality of the fit, the fitting results from the above two approaches suggest that the $T_2$ transition in low field (< 4 T) is consistent with a band gap of $\Delta \approx 38 \pm 4$ meV, similar to the value from fitting the $T_1$ transition. From this gap value, we obtained an effective Fermi velocity $v_F^{T2} \approx (1.20 \pm 0.01) \times 10^6$ m s$^{-1}$ for the $T_2$ transition.

In the least squares fit for the $T_3$ transition based on equation (5) in the main text, we define the sum of squared residuals in the approaches (1) and (2) described above as $\chi_1^2$ and $\chi_2^2$, respectively. If we assume $\Delta$ and $v_F$ are field-independent, we find $\chi_2^2 = 16\chi_1^2$ from the fit as shown in Supplementary Figure 5, which suggests that the second approach with $\Delta = 38 \pm 4$ meV is not a good fit to the data. Instead, the $T_3$ transition is much better described by the first approach, which yields $\Delta' = 96 \pm 3$ meV and $v_F^{T3'} \approx (1.12 \pm 0.01) \times 10^6$ m s$^{-1}$. The significantly larger gap value $\Delta'$ obtained this way compared to $\Delta = 38 \pm 4$ meV from $T_1$ and $T_2$ transitions strongly suggests a deviation of the $T_3$ transition from behaviors expected within the single particle picture. Therefore, we use a *phenomenological* approach to discuss the $T_3$ transition: it is natural to assume a similar gap value of $38 \pm 4$ meV for the $T_3$ transition, corresponding to the zero-field gap in equation (5) in the main text, and attribute all deviations of $T_3$ transition from behaviors described by equation (5) to an effective field-dependent parameter $v_F^{T3}(B)$. The *phenomenological* description employed here assumes that the gap associated with the $T_3$ transition is similar to those obtained from $T_1$ and $T_2$ transitions, all of which are corresponding to the zero field gap of the gapped graphene. For each data points of the $T_3$ transition in figure 3 of the main text, we calculate the corresponding $v_F^{T3}$ value based on equation (5) using $\Delta = 38 \pm 4$ meV, and the resulting $v_F^{T3}(B)$ is shown in figure 4b of the main text in the form of $v_F^{T3}(B)/v_F^{T1}$. This description can illustrate the main features of $T_3$ transition that are distinct from those characterized by equation (5) in the main text.

**Supplementary Note 5 | Discussions on interaction effects in pristine graphene and data for CVD graphene on SiO$_2$.**

It is imperative to go beyond the single particle picture and examine whether our results can be explained by interaction effects in pristine graphene without invoking the notion of a band gap. Our magneto-transmission experiments probe transitions between LLs that involve exciting an electron in an occupied LL n to an unoccupied LL n', leaving behind a hole in the initial LL n. Therefore, electron-hole excitations between LLs including e-e interactions[12-15] should be considered. The effects in pristine graphene have been studied theoretically, taking into account contributions such as direct Coulomb interactions between the excited electrons and holes (exciton binding energy) and the exchange self-energy of electrons and holes[12-14]. It is found that these interactions lead to corrections $\Delta E_{n,n'}$ to the non-interacting LL transition energy $E_{n,n'}$, and the corrections $\Delta E_{n,n'}$ all exhibit $\sqrt{B}$ field dependence[12-14]. Consequently, the LL transition energies renormalized by e-e interactions $E_{n,n'} + \Delta E_{n,n'}$ still scale with $\sqrt{B}$ but are characterized by a renormalized and field-independent Fermi velocity $v_F^*$, which is different from the non-interacting value $v_F$. Moreover, because the magnitude of the many-body effects on the electron-hole excitations is intimately related to both the initial and excited LLs, the renormalized velocity varies for different inter-LL transitions. It is shown theoretically[12,13,15] that the renormalized Fermi velocity values $v_F^*$ for transitions $T_2$ and $T_3$ are larger than that for $T_1$ tramsition. The experimental signature of this prediction is that the LL transition energy ratios $E_{T2}/[(\sqrt{2} + 1)E_{T1}]$ and $E_{T3}/[(\sqrt{3} + \sqrt{2})E_{T1}]$ are higher than one; the latter value is the result from single particle picture.

To investigate the effect of many-body interactions on the LL transitions in pristine graphene, we carried out magneto-transmission measurements on graphene samples that are CVD grown on copper and transferred to SiO$_2$/Si substrates. The Fermi energy E$_F$ of these samples is around 160 meV, which is determined from cyclotron resonance measured in the far-infrared region[16]. We observed T$_1$, T$_2$ and T$_3$ transitions (as discussed in the main text) in these samples, as shown in Supplementary Figure 6a. The dip features below the T$_1$ transition in Supplementary Figure 6a are due to intraband LL transitions, which will be reported in a future publication. The LL transition energies are shown in Supplementary Figure 6b, all of which show linear $\sqrt{B}$ field dependence and, importantly, converge to zero energy in zero field as described by equation (1) in the main text. These results are consistent with experimental study of mechanically exfoliated single layer graphene[5]. Moreover, we plot LL transition energy ratios $E_{T2}/[(\sqrt{2} + 1)E_{T1}]$ and $E_{T3}/[(\sqrt{3} + \sqrt{2})E_{T1}]$ as a function of magnetic field as shown in figure 4a of the main text. Within the single particle picture, all the LL energies can be described by a single non-interacting value of $v_F$, so these two energy ratios should be one and field-independent. However, the two ratios for our data on graphene on SiO$_2$ are both higher than one in the field range where these two transitions were observed, which is consistent with the prediction of many-body theory. Similar results were firstly reported in infrared measurements on exfoliated graphene samples on SiO$_2$[5]. Therefore, our measurements together with previous experiments[5] have corroborated the theoretical predictions on many body effects in pristine graphene[12-15].

Our data on graphene/h-BN cannot be explained by interacting massless Dirac fermions in pristine graphene. Firstly, the LL transition energies of interacting massless Dirac fermions scale with $\sqrt{B}$, which cannot describe the observed non-zero energy intercepts at zero magnetic field for our graphene/h-BN samples. Moreover, the LL transition energy ratios of graphene on h-BN show very different behaviors. Figure 4a of the main text depicts $E_{T2}/[(\sqrt{2} + 1)E_{T1}]$ and $E_{T3}/[(\sqrt{3} + \sqrt{2})E_{T1}]$ as a function of magnetic field for our data on graphene/h-BN, both of which are lower than one. This behavior is in direct contrast to the established behaviors of interacting massless Dirac fermions described above. We also note that electron-phonon interactions of massless Dirac fermions cannot lead to LL transition behaviors observed here[17]. These analyses clearly demonstrate that our data on graphene/h-BN cannot be described by interacting massless Dirac fermions in pristine graphene.

If we define a critical field B$_C$ above which the E$_F$ is between the zero-th LL and first LL, we find that $B_C$ ~ 18 T for our CDV samples on SiO$_2$ with $E_F$ ~ 160 meV based on equation (1) in the main text. So the data for graphene on SiO$_2$ in figure 4a of the main text and Supplementary Figure 6 are in the field range of $B_C$ to 2$B_C$. For graphene on h-BN samples, we observe T$_1$ transition staring at ~ 2 T, which indicates $B_C$ is below 2 T. Thus, figure 4a in the main text shows the data for graphene on h-BN in the field range of $B_C$ to 5$B_C$. Therefore, the comparison between data for graphene on SiO$_2$ and graphene/h-BN shown in figure 4a of the main text is appropriate. Also, previous IR study[5] of LLs in exfoliated graphene on SiO$_2$ was performed by changing the gate voltage to ensure that the data is taken above $B_C$ at all magnetic fields.

Previous magneto-optical experiments on LL transitions in epitaxial graphene on SiC all showed results consistent with single particle picture, namely, $T_1$, $T_2$ and $T_3$ transitions can be described with a single value of $v_F$[18-20]. However, in those studies of graphene on SiC LL transitions such as $T_1$, $T_2$ and $T_3$ were only observed in the experiments carried out on *multilayer* graphene with random twist angles between graphene layers. It is beyond the scope of this study to discuss why LL transitions of *multilayer* graphene on SiC can be described by a single particle picture, but we believe that the coupling between graphene layers are very likely to significantly modify the behaviors of LL transitions in multilayer graphene compared to the intrinsic properties of single layer graphene. We stress that the LL transition energy ratios $E_{T2}/[(\sqrt{2} + 1)E_{T1}]$ and $E_{T3}/[(\sqrt{3} + \sqrt{2}) E_{T1}]$ are experimentally demonstrated to be greater than one in *single layer* graphene, indicating many body effects of massless Dirac fermions in pristine graphene, which is observed in both CVD grown graphene in our study and exfoliated single layer samples in previous IR studies[5].

## Supplementary References